\documentclass[twocolumn,prl,amsmath,amssymb]{revtex4-1}

\usepackage{graphicx}
\usepackage{dcolumn}
\usepackage{bm}
\usepackage{amsmath}
\usepackage{comment}
\usepackage{xcolor} 
\usepackage{siunitx}

\usepackage{hyperref}

\begin{document}

\preprint{APS/123-QED}

\title{Scalable and Robust Photonic Integrated Unitary Converter}

\author{Ryota Tanomura}
\author{Rui Tang}%

\author{Toshikazu Umezaki}%

\author{Go Soma}%

\author{Takuo Tanemura}%
\author{Yoshiaki Nakano}
\affiliation{%
 Department of Electrical Engineering and Information Systems, School of Engineering,\\ The University of Tokyo, 7-3-1 Hongo, Bunkyo-ku, Tokyo, 113-8656, Japan
}%

\date{\today}

\begin{abstract}
Optical unitary converter (OUC) that can convert a set of {\it N} mutually orthogonal optical modes into another set of arbitrary {\it N} orthogonal modes is expected to be the key device in diverse applications, including the optical communication, deep learning, and quantum computing. While various types of OUC have been demonstrated on photonic integration platforms, its sensitivity against a slight deviation in the waveguide dimension has been the crucial issue in scaling {\it N}. Here, we demonstrate that an OUC based on the concept of multi-plane light conversion (MPLC) shows outstanding robustness against waveguide deviations. Moreover, it becomes more and more insensitive to fabrication errors as we increase {\it N}, which is in clear contrast to the conventional OUC architecture, composed of 2$\times$2 Mach-Zehnder interferometers. The physical origin behind this unique robustness and scalability is studied by considering a generalized OUC configuration. As a result, we reveal that the number of coupled modes in each stage plays an essential role in determining the sensitivity of the entire OUC. The maximal robustness is attained when all-to-all-coupled interferometers are employed, which are naturally implemented in MPLC-OUC.
\end{abstract}

\maketitle


{\it{Introduction}-} 
Reconfigurable unitary conversion in the optical domain is a crucial operation in various photonic signal-processing applications, both in the classical and quantum systems.
In particular, a large-scale integrated optical unitary converter (OUC) would be the key device in realizing low-power all-optical multi-input-multi-output processors for mode-multiplexed optical communication  \cite{SDM1,SDM2,SDM3_tang}, optical accelerators for deep learning \cite{ONN1,ONN2,ONN3,ONN4,ONN5}, and linear-optical circuits for quantum computing \cite{quantum1,quantum2,quantum3,quantum_PRA,quantum4,quantum5,quantum6}.\par
An integrated {\it N}$\times${\it N} OUC can be constructed by cascading multiple stages of 2$\times$2 OUCs in a mesh configuration \cite{Reck,Miller,Clements}. 
Each 2$\times$2 OUC corresponds to a Mach-Zehnder interferometer (MZI), which comprises two 50:50 directional couplers (DCs) and two phase shifters.
By setting all $N^2$ phase shifters appropriately, arbitrary {\it N}$\times${\it N} unitary operation can be obtained in a reconfigurable manner \cite{Reck,Clements}.

\par
In such MZI-based OUCs, however, the accuracy of {\it N}$\times${\it N} unitary operation depends sensitively on the deviations of waveguide dimensions, especially when {\it N} is large \cite{Miller2, Pai, ONN4, Saygin}. 
This issue is related to the fact that the optical interactions in the respective MZIs are inherently local, which can easily result in a sparse {\it N}$\times${\it N} matrix when they deviate from ideal conditions. 
We should note that such property is rather a general characteristic observed in other analog and quantum computing systems as well, such as the coherent Ising machines, quantum annealers, and multiqubit processors 
{\cite{Ising_all3,Ising_all,Ising_all2, quantum_processor,quantum_processor2}},
where the dense connectivity among multiple nodes has a crucial impact on the performance of the entire system.
To cope with this issue of MZI-OUC, it is proposed to insert extra redundant phase shifters to ensure the necessary full connectivity \cite{Miller2, Pai}, at the cost of increased complexity, footprint, and power consumption.
Various global optimization techniques may also be employed to compensate for the device imperfectness \cite{ONN2, Pai, optimization}, which, however, requires the errors to be within some acceptable range \cite{Pai}.
\par
As a fundamentally different approach as compared with MZI-OUC, an OUC based on the multi-plane light conversion (MPLC) concept has been studied recently \cite{MPLC1,MPLC2,tanomura1,Saygin}. 
In this scheme, arbitrary {\it N}$\times${\it N} unitary  conversion is obtained through cascaded stages of an {\it N}-dimensional fixed dense unitary transformation (i.e., mode-mixing layer) and a phase shifter array. MPLC-OUCs have been implemented successfully by free-space optics \cite{MPLC2,MPLC3,MPLC4,space_MPLC} and on integrated photonic platforms \cite{SDM3_tang,tanomura1,tanomura_OE}. Due to the mode-mixing layer, which ensures all-to-all coupling via an {\it N}$\times${\it N} dense matrix, MPLC-OUC is expected to exhibit inherently different scalability and robustness, compared with the conventional MZI-OUC.
\par
In this letter, we reveal that the MPLC-OUC with multiport DCs has unique scalability and excellent robustness against waveguide deviations without the need for redundant phase shifters. Surprisingly, the error sensitivity of the MPLC-OUC drops rapidly at large {\it N}. This is in clear contrast to the MZI-OUC, having the same number of phase shifters, which becomes more and more sensitive as {\it N} increases.
By considering generalized circuit configurations with various degrees of coupling, we comprehensively evaluate the physical origin of the difference between the two schemes. As a result, we find that the number of coupled modes at each mode-mixing layer plays an essential role in determining the robustness of the entire OUC; the maximal robustness is obtained when all-to-all-coupled interferometers are employed, which are naturally implemented in MPLC-OUC.

{\it{Principle -}}
An OUC converts an $N$-dimensional vector $\bf a^{in}$, which describes the complex amplitudes of $N$ optical modes at the input, into another $N$-dimensional vector $\bf a^{out}$ at the output. Here, $\bf a^{out}$ can be expressed as ${\bf a^{out}}={\bf Ua^{in}}$, where $\bf U$ is an {\it N}$\times${\it N} unitary matrix.
\par
\begin{figure}[t]
\includegraphics[width=\linewidth]{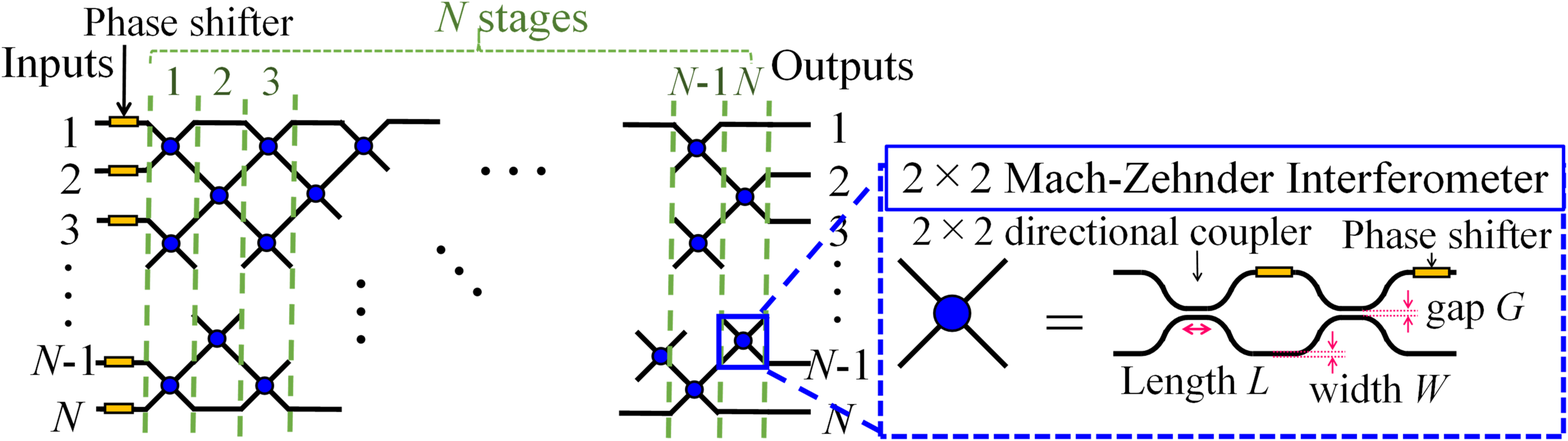} \\
(a)
\includegraphics[width=\linewidth]{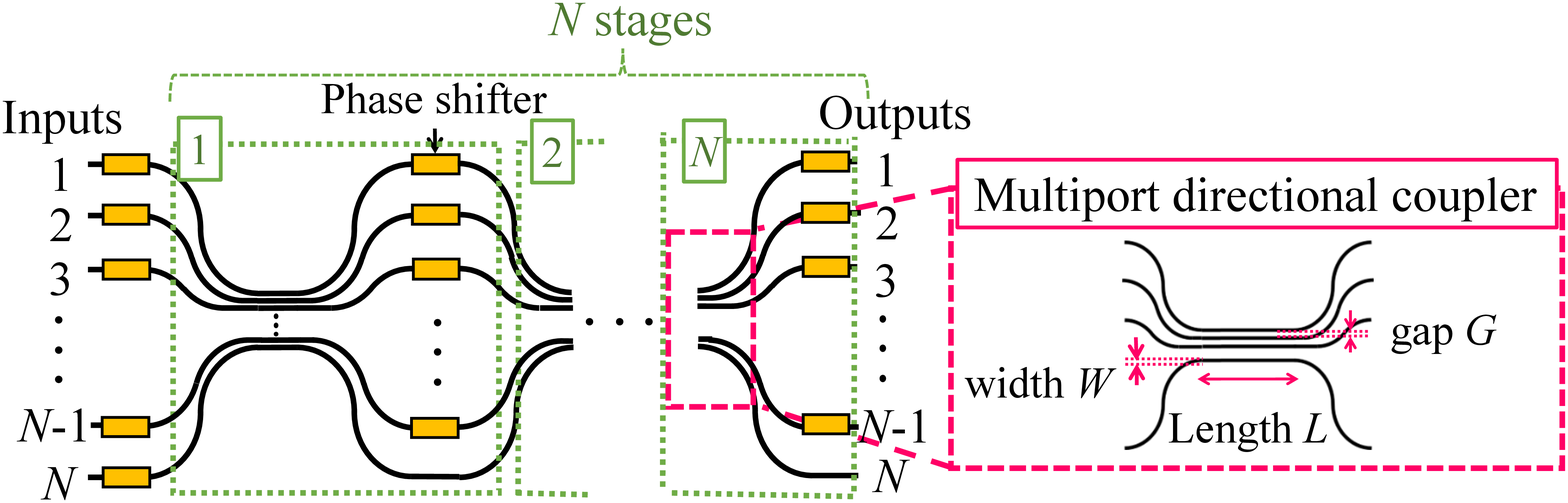} \\
(b)
\caption{\label{fig:1} Schematics of (a) MZI-OUC and (b) MPLC-OUC with multiport DC.}
\end{figure}

Figure \ref{fig:1}(a) shows the typical architecture of the MZI-OUC \cite{Clements}. It consists of 2$\times$2 MZIs, each of which comprises two 2$\times$2 DCs and two tunable phase shifters as shown in the inset of Figure \ref{fig:1}(a). The splitting ratio of the DC needs to be 50:50.
Since the number of MZIs is {\it N}({\it N}-1)/2 and {\it N} additional phase shifters are required at the input stage, the total number of phase shifters is $N^2$.
By tuning all $N^2$ phase shifters, arbitrary  $\bf U$ is obtained \cite{Clements}.
\par
In contrast, Fig. \ref{fig:1}(b) shows the architecture of the integrated MPLC-OUC with multiport DCs \cite{Tang_mDC, tanomura1}. It consists of {\it N} stages of an {\it N}-port DC that functions as a mode-mixing layer and an {\it N}-port phase shifter array that can be controlled independently.
The transfer matrix of the entire circuit is written as
\begin{eqnarray}
{\bf U} = {\bf \Phi^{(\it N)}}{\bf M^{(\it N)}}{\bf \Phi^{(\it N-1)}}\cdots{\bf \Phi}^{(1)}{\bf M}^{(1)}{\bf \Phi}^{(0)},
\end{eqnarray}
where ${\bf M}^{(j)}$ and ${\bf \Phi}^{(j)}$ ($j = 1, 2, ..., N$) are the {\it N}$\times${\it N} transfer matrices of the multiport DC and the phase shifter array, respectively, at the $j$-th stage.
${\bf \Phi}^{(j)}$ is a diagonal matrix, expressed as
\begin{eqnarray}
{\bf \Phi}^{(j)} = \rm{diag}[{\exp}( i {\bm \phi}^{({\it j})})].
\end{eqnarray}
Here, $\bm{\phi}^{(j)} = (\phi_1^{(j)}, \phi_2^{(j)}, \cdots, \phi_{\it N}^{(j)})$, where $\phi_m^{(j)}$ is the phase shift at the $m$-th port of the $j$-th stage.
Note that we need {\it N} phase shifters at the input stage ($j = 0$), while the other stages require only $N-1$ phase shifters.
The total number of phase shifters is, therefore, $ N+N(N-1) = N^2$, which is identical to MZI-OUC.
As a unique feature of MPLC-OUC in contrast to MZI-OUC, the mode-mixing layers ${\bf M}^{(j)}$ neither need to be identical in all stages nor uniform among all ports, but are only requested to provide substantial coupling among all modes \cite{Tang_mDC,Saygin}.

{\it Scalability analysis -} 
The robustness and scalability of the two OUC architectures presented in Fig. \ref{fig:1} are compared numerically. As a test case of interest, we consider standard silicon photonic circuits with a 220-nm-thick silicon-on-insulator (SOI) device layer. The waveguide width {\it W} and gap {\it G} at the DC sections (see Fig. \ref{fig:1} for definitions) are set to 460 nm and 250 nm, respectively. The 2$\times$2 DC length in Fig. \ref{fig:1}(a) is then fixed to 20 $\si{\micro}$m, corresponding to the 50:50 splitting length under this condition. In contrast, the coupling length of the {\it N}-port DC in Fig. \ref{fig:1}(b) is scaled as {\it N} $\times$ 10 $\si{\micro}$m. In both cases, the transfer matrix of the DC is derived numerically by the eigenmode expansion method (EME). On the other hand, we assume an ideal lossless phase shifter, each driven independently by an 8-bit digital-to-analog converter (DAC). 
\par
First, the sensitivity of the OUC performance is investigated against variations in {\it W} and {\it G}. 
For each case, 
all $N^2$ phase shifters are optimized to obtain the target unitary matrix ${\bf U'}$.
Then, the deviation of the generated {\it N}$\times${\it N} transfer matrix {\bf U} from ${\bf U'}$ is evaluated by calculating the mean-square error (MSE) $f_{\rm{MSE}}$, defined as
\begin{eqnarray}
f_{\rm{MSE}} =  \frac{1}{{\it N}^2}\sum_{i,j}^{N} |{U}_{ij}'-{U}_{ij}|^2.
 \end{eqnarray}
To ensure the reliability of the analysis, we generate 20 different Haar random {\it N}$\times${\it N} matrixes as the target unitary matrix ${\bf U'}$ and take the average of $f_{\rm{MSE}}$ in all cases of ${\bf U'}$.
In optimizing MZI-OUC, we first calculate the 2$\times$2 transfer matrix of an MZI for respective cases of $W$ and $G$, and then derive the optimal conditions of all phase shifters for each target matrix ${\bf U'}$ through the decomposition algorithm \cite{Clements}.
It is confirmed that the resultant performance of OUC is consistent with the best case obtained through the simulated annealing optimization. In optimizing MPLC-OUC, the simulated annealing algorithm \cite{SA} is employed when $N \le 32$, whereas the adaptive moment estimation (ADAM) algorithm \cite{ADAM} is applied when {\it N} = 128 to derive the quasi-optimal conditions of phase shifters. Note that these general-purpose algorithms are employed for the sake of simplicity in this work, but a more efficient algorithm specific to MPLC-OUC may be developed in future to accelerate the convergence.
\begin{figure*}[t]
\includegraphics[width=\linewidth]{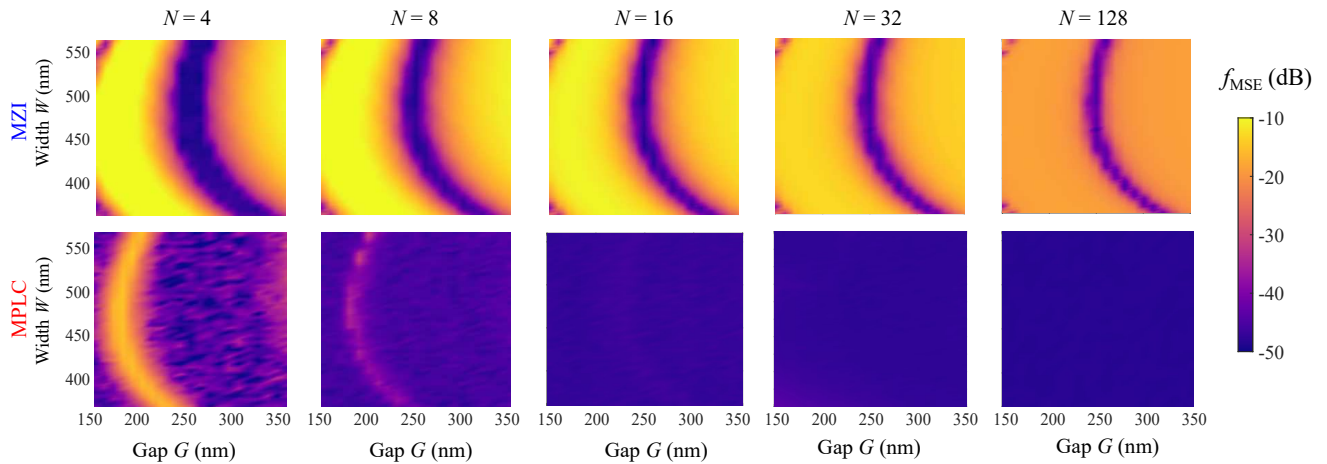}
\caption{\label{fig:2} $f_{\rm{MSE}}$ of the obtained ${\bf U}$ after tuning all phase shifters as a function of width {\it W} and gap {\it G} for MZI-OUCs (upper row) and MPLC-OUCs (lower row) with {\it N} = 4, 8, 16, 32, and 128.}
\end{figure*}
\begin{figure}
\includegraphics[width=0.8\linewidth]{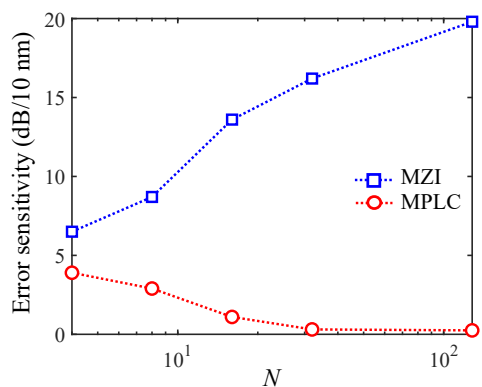}
\caption{\label{fig:3} Error sensitivities (defined as increase in $f_{\rm{MSE}}$ when $W$ or $G$ deviates by 10 nm) of MZI-OUC and MPLC-OUC versus $N$, derived at the optimal points in Fig. \ref{fig:2}. }
\end{figure}
\par
Figure \ref{fig:2} shows $f_{\rm{MSE}}$ after optimizing all $N^2$ phase shifters in both architectures with increasing $N$. In all cases, the minimum $f_{\rm{MSE}}$ is limited to around -50 dB due to the 8-bit resolution of the DACs. 
For the case of MZI-OUC, $f_{\rm{MSE}}$ is suppressed below -45 dB only within a limited range of {\it W} and {\it G}, at which each 2$\times$2 DC attains a precise 50:50 splitting ratio.
Moreover, this regime shrinks rapidly as we increase {\it N}, meaning that arbitrary unitary conversion can no longer be obtained even after global optimization of all $N^2$ phase shifters.
This issue would severely limit the scalability of this architecture.
In contrast, for the case of MPLC-OUC, Fig. \ref{fig:2} clearly shows that $f_{\rm{MSE}}$ is less sensitive to {\it W} and {\it G}. More interestingly, this trend is enhanced as {\it N} increases to 128, where $f_{\rm{MSE}}$ is nearly insensitive to {\it W} and {\it G}.
\par
For quantitative evaluation, the error sensitivities of both OUC configurations are plotted in Fig. \ref{fig:3} as a function of {\it N}. Here, the error sensitivity is defined as the largest increase in $f_{\rm{MSE}}$ when {\it G} or {\it W} deviates by 10 nm from the optimum point in Fig. \ref{fig:2}. Figure \ref{fig:3} clearly indicates that MZI-OUC and MPLC-OUC show contrary trends; the error sensitivity of MZI-OUC increases monotonically with $N$ and approaches 20 dB/(10 nm) at $N = 128$, whereas that of MPLC-OUC converges to 0 as we increase $N$. Given that the state-of-art silicon-photonic waveguides inevitably have nanometer-scale errors \cite{fabrication_1, fabrication_2}, this unique robustness of MPLC-OUC architecture would provide a significant advantage in constructing large-scale OUCs with $N > 100$.

%

\begin{figure}[t]
\includegraphics[width=1.0\linewidth]{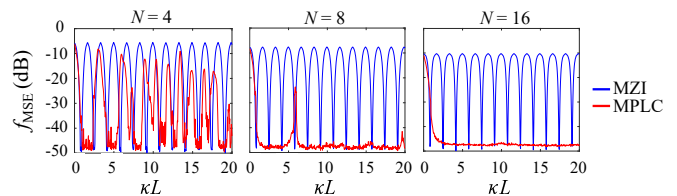}
\caption{\label{fig:4} $f_{MSE}$ as a function of {\it ${\kappa}L$}
for the MZI-OUC and MPLC-OUC with $N$ = 4, 8, and 16.}
\end{figure}

The results shown in Fig. \ref{fig:2} and \ref{fig:3} imply that the accuracy of MZI-OUC depends sensitively on the splitting ratio of each 2$\times$2 DC, whereas MPLC-OUC seems rather insensitive to the precise characteristic of the multiport DC, especially when $N$ increases.
To verify this assumption, we derive $f_{\rm{MSE}}$ of two schemes as a function of {\it $\kappa$L}, where {\it $\kappa$} is the coupling coefficient between adjacent waveguides at DCs and {\it L} is the coupling length.
The results are shown in Fig. \ref{fig:4}. 
We can confirm that the performance of MZI-OUC changes periodically with {\it $\kappa$L}, in accordance with the splitting ratio. 
On the contrary, MPLC-OUC becomes more and more robust against the change in the DC characteristic as $N$ increases, in agreement with Fig.3.

\begin{figure}[t]
\includegraphics[width=\linewidth]{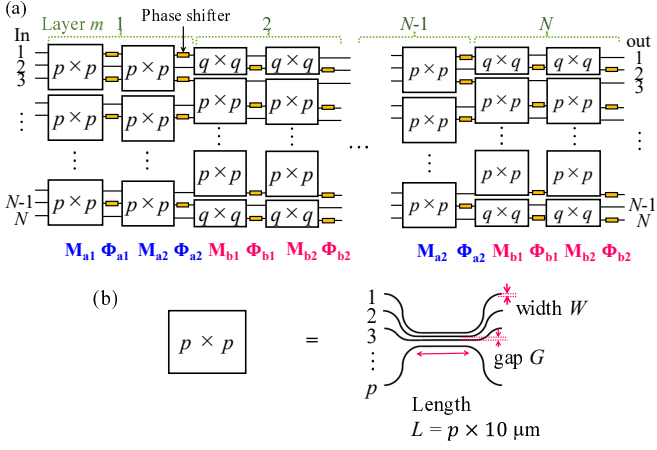}
\caption{\label{fig:5} Generalized {\it N}-port OUC with a different degree of coupling, represented by $p$. (a) Overall circuit configuration, and (b) each ${\it p}\times{\it p}$ mode-mixing block, realized by a {\it p}-port directional coupler.
The MZI-OUC and MPLC-OUC correspond to the cases with (${\it p}$, ${\it q}$) = (2, 1) and ($N$, 0), respectively.
}
\end{figure}

{\it Effects of the all-to-all coupling to error sensitivity -}
From the above results and discussions, we assume that the all-to-all coupling at each stage of MPLC-OUC plays an essential role in attaining the unique scalability and excellent error tolerance.
To investigate this aspect in a comprehensive manner, we consider a generalized OUC configuration as shown in Fig. \ref{fig:5}(a).
The transfer matrix 
of this circuit is expressed as 
\begin{eqnarray}
{\bf U} =&& {\bf T}^{({\it N})}{\bf T}^{({\it N}-1)}\cdots{\bf T}^{({1})},
\end{eqnarray}
where ${\bf T}^{({\it i})}$ is the transfer matrix of the {\it i}-th layer and is written as
\begin{eqnarray}
{\bf T}^{({\it i})} = \left\{\begin{array}{ll}  {\bf {\Phi}}_{\bf b2}^{(\it i)}\cdot{\bf M}_{\bf b2}^{(\it i)}\cdot{\bf \Phi}_{\bf b1}^{(\it i)}\cdot{\bf M}_{\bf b1}^{(\it i)}  \text{ (when {\it i} is odd)}, \\ [+5pt]
{\bf {\Phi}}_{\bf a2}^{(\it i)}\cdot{\bf M}_{\bf a2}^{(\it i)}\cdot{\bf \Phi}_{\bf a1}^{(\it i)}\cdot{\bf M}_{\bf a1}^{(\it i)}  \text{ (when {\it i} is even)}.
\end{array}
\right.
\end{eqnarray}
Here, ${\bf \Phi_a} ({\bf \Phi_b})$ describes each phase shifter stage, where the phase shifters are installed only at the odd (even) ports as shown in Fig. \ref{fig:5}(a). On the other hand, ${\bf M_a}$ and ${\bf M_b}$ represent the mode-mixing stages. As shown in Fig. \ref{fig:5}(a), ${\bf M_a}$ consists of an array of ${\it p}\times{\it p}$ multiport DC, described in Fig. 5(b). In contrast, ${\bf M_b}$ consists of two blocks of ${\it q}\times{\it q}$ mode-mixing matrixes at the top and the bottom channels in Fig. 5(a), whereas other ports are connected to ${\it p}\times{\it p}$ matrixes.
Note that $p$ needs to be a submultiple of $N$.
On the other hand, $q$ is either $p/2$ or 0.

Using this generalized model, various types of OUC with different degrees of mode coupling can be analyzed in a unified manner. For example, the MZI-OUC shown in Fig. \ref{fig:1}(a) corresponds to a case with (${\it p}$, ${\it q}$) = (2, 1). By increasing $p$, we can increase the degree of coupling at each stage. 
As an extreme case, a circuit equivalent to the MPLC-OUC in Fig. \ref{fig:1}(b) is represented as (${\it p}$, ${\it q}$) = (${\it N}$, 0).
%
Through the same procedure as Fig. \ref{fig:2}, $f_{\rm{MSE}}$ is obtained numerically as a function of $W$ and $G$ to investigate the dependence of the error sensitivity on $(p, q)$.

\begin{figure}[t]
\includegraphics[width=0.98\linewidth]{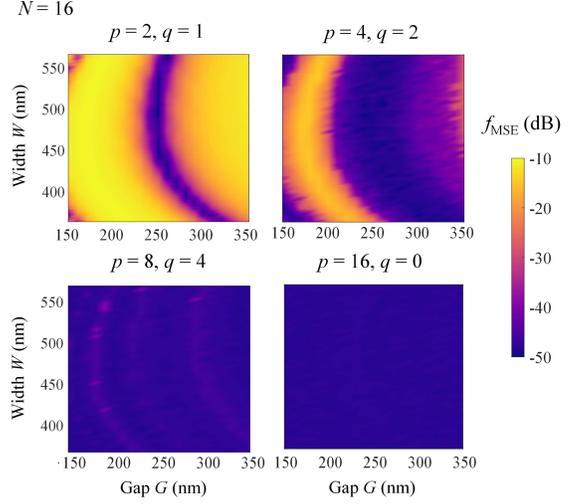}
\caption{\label{fig:6} $f_{\rm{MSE}}$ of 16$\times$16 OUC ({\it N} = 16) with different ({\it p}, {\it q})  as a function of width ({\it W}) and gap ({\it G}) of each DC. 
As {\it p} increases, $f_{\rm{MSE}}$ becomes more and more insensitive to the deviations.}
\end{figure}
Figure \ref{fig:5} shows $f_{\rm{MSE}}$ of 16$\times$16 OUC with various (${\it p}$, ${\it q}$).
We can clearly see that as ${\it p}$ increases, $f_{\rm{MSE}}$ becomes more and more insensitive against errors.
In particular, in the extreme case of (${\it p}, {\it q}$) = (16, 0), the MSE is kept below -40 dB for all values of {\it W} and {\it G}. We, therefore, attribute the unique robustness of the MPLC-OUC to the all-to-all coupling at each mode-mixing stage, which ensures the necessary full connectivity of all modes.

{\it Conclusion-} We have investigated the scalability and error tolerance of different types of OUCs.
Unlike the conventional OUC architectures, composed of 2$\times$2 MZIs, MPLC-OUC with multiport DCs was demonstrated to show superior robustness against waveguide deviations. Moreover, we found that it becomes more and more robust as the port count {\it N} increases, which is in clear contrast to MZI-OUC, whose sensitivity increases with {\it N}.

From comprehensive analyses on generalized OUC configurations, we revealed that the all-to-all coupling inside each multiport DC in MPLC-OUC plays an essential role in enhancing the robustness and provides the crucial difference from MZI-OUC. More specifically, the error sensitivity of MZI-OUC originates from the localized interactions at MZIs, which may result in a sparse $N\times N$ matrix at large {\it N} when they deviate from ideal conditions. In contrast, the MPLC-OUC naturally ensures the necessary full connectivity among all modes without the need for redundant phase shifters. 
The physical insight provided in this work should be useful in building large-scale OUCs for diverse applications.

\bigskip
We thank Yoshihiko Hasegawa, Quoc Hoan Tran, Tan Van Vu, and Taichiro Fukui for the fruitful discussions. This work was in part supported by the JSPS KAKENHI Grant Number JP20J21861 and JP26000010.



\end{document}